\documentclass{aa}
\usepackage{graphicx}
\usepackage{enumerate}
\usepackage{txfonts}
\usepackage{natbib}
\usepackage{color}

\newcommand{\msun}{\ensuremath{\mathrm{M}_\odot}}

\bibpunct{(}{)}{;}{a}{}{,}

\begin{document}

\title{Explaining the Ba, Y, Sr, and Eu abundance scatter in metal-poor halo stars: constraints to the r-process}

\author {G. Cescutti\inst{1} \thanks {email to: cescutti@aip.de} \and
  C. Chiappini \inst{1} }

\institute{Leibniz-Institut f\"ur Astrophysik Potsdam, An der
  Sternwarte 16, 14482, Potsdam, Germany }

\date{Received xxxx / Accepted xxxx}

\abstract {Thanks to the heroic observational campaigns carried out in
  recent years we now have large samples of metal-poor stars for
  which measurements of detailed abundances exist. In particular,
  large samples of stars with metallicities $-$5 $< $ [Fe/H] $ < -$1
  and measured abundances of Sr, Ba, Y, and Eu are now available. These
  data hold important clues on the nature of the contribution of
    the first stellar generations to the enrichment of our
  Galaxy.}{We aim to explain the scatter in Sr, Ba, Y, and Eu
  abundance ratio diagrams unveiled by the metal-poor halo stars.}
{We computed inhomogeneous chemical evolution models for the Galactic
  halo assuming different scenarios for the r-process site:
  the electron-capture supernovae (EC) and the
  magnetorotationally driven (MRD) supernovae scenario. We also
  considered models with and without the contribution of fast-rotating
  massive stars (spinstars) to an early enrichment by the s-process. A
  detailed comparison with the now large sample of stars with measured
  abundances of Sr, Ba, Y, Eu, and Fe is provided (both in terms of
  scatter plots and number distributions for several abundance
  ratios).}  {The scatter observed  in these abundance ratios of
  the very metal-poor stars (with [Fe/H] $<-$2.5) can be explained by
  combining the s-process production in spinstars, and the r-process
  contribution coming from massive stars. For the r-process we have
  developed models for both the EC and the MRD scenario that
   match the observations.}{With the present observational and
  theoretical constraints we cannot distinguish between the EC and the
  MRD scenario in the Galactic halo. Independently of the r-process
  scenarios adopted,  the production of elements by an s-process
    in spinstars is needed to reproduce the spread in 
    abundances of the light neutron capture elements (Sr and Y) over
  heavy neutron capture elements (Ba and Eu). We provide a way
  to test our suggestions by means of the distribution of the Ba
  isotopic ratios in a [Ba/Fe] or [Sr/Ba] vs. [Fe/H] diagram.}


\titlerunning{r-processes in the halo}

\authorrunning{Cescutti \& Chiappini}

\maketitle

\section{Introduction}

The site for the production of the heaviest elements built via rapid
neutron captures (the so-called r-process) is still unclear, and has
been driving large theoretical efforts
\citep[e.g.][]{Goriely13,Nakamura13, Wanajo13, Qian12, Winteler12,
  Arcones11, Thielemann11}.  The r-process requires high neutron
  fluxes (i.e., the high neutron-to-seed ratios needed for the
  r-process to occur). The site of the r-process must also
  reproduce the abundance patterns seen in strongly r-process-enhanced 
metal-poor stars (which match the solar r-process pattern in
  a wide range of elements), and hence enrich the interstellar medium (ISM)
 on short timescales.

In our latest work \citep{Cescutti13}, we studied the impact on the
chemical evolution of the Galactic halo of the s-process 
generated by massive fast-rotating metal-poor stars (spinstars). We
showed that spinstars can explain the long-standing
problem of the [Sr/Ba] spread in the Galactic halo \citep[for
alternative scenarios see][]{ArcoMonte11,AokiSuda13}. However, to
achieve this, it was necessary to  consider the contribution of an
  r-process to the chemical enrichment. In \citet{Cescutti13}, we
followed the scenario described by \citet{WNJ09}, where the r-process
occurs in a relative narrow mass range (8-10~\msun). We underline that
 these assumptions on the r-process did not influence our main
result,  which was to show how spinstars can explain the Sr/Ba ratios.

In the present work we verify this by testing other r-process
scenarios.  First, this offers the opportunity to confirm the
important role of spinstars not only in the chemical evolution of the
light elements such as C and N \citep{Chiappini06,Chiappini08}, but
also for the heavier elements \citep{Pigna08, Chiappini11,
  Frisch12}. Second, we aim to find observational constraints on the
nature of the r-process by studying chemical evolution models of the
earliest phases of the chemical enrichment of our Galaxy.

In the present work we compute a new chemical evolution model 
  that includes the site of production of r-process recently suggested
by \citet{Winteler12}. These authors suggested that
magnetorotationally driven supernovae might be the source of the
r-process in the early Galaxy. These SN explode in a rare progenitor
configuration that is characterized by a high rotation rate and a strong
magnetic field necessary for the formation of bipolar jets.  The
findings of \citet{Winteler12} suggest that the second and third peaks
of the solar r-process distribution can be reproduced well. Here, we
test whether this site for the r-process provides an enrichment for
the earliest phases of the Galactic chemical evolution 
  consistent with the abundances observed in metal-poor halo stars.

We anticipate that the results we obtain in the Galactic halo for
the Winteler scenario cannot clearly be distinguished from the r-process
scenario based on electron-capture SNe used in \citet{Cescutti13}, at
least not before a substantial improvement in the number of stars
measured in the Galactic halo  has provided stronger constraints.
Therefore, spinstars play a key role in this scenario as well, and the
oldest halo stars are formed from an ISM enriched by both  r- and
s-processes.

A clear prediction of both models is that EMP stars with a high [Sr/Ba]
ratio should be almost entirely enriched by the s-process. This prediction
is original and differs from the other possible scenarios in which the
spread in [Sr/Ba] ratio is explained  by a weak
r-process \citep{ArcoMonte11} or a truncated r-process
\citep{AokiSuda13}. A way to distinguish in this mixture between s-process
and r-process in the early phase of the Galaxy formation is to examine
the prediction of our models for the Ba isotopes.  

The s-process preferentially produces even isotopes, whereas the
r-process produces  approximately the same amount of odd and even
isotopes. According to nucleosynthesis calculations \citep{Arla99}, it
is expected  that an odd fraction of Ba isotopes ($f_{odd}$) = 0.11 $\pm$
0.01  occurs in the case of a pure s-process, and an {\it fodd} = 0.46 $\pm$
0.06 in the case of pure r-process. \citet{Magain1995} measured for
the first time the isotopic ratio of a very bright halo star, HD
140283,  finding an s-process signature ([Sr/Ba]=0.9), which 
agrees with our theoretical results.  However, his results have
been challenged and still need to be confirmed
\citep{Lambert2002,Collet2009,Gallagher10}.  The biggest challenge is
to correctly  take into account the 3D effects on the line
formation. More recently, \citet{Gallagher12} have again attempted to
measure isotopic ratios in other metal-poor stars, but all their
candidates are expected to be s-process dominated.  Although the
measurement of the Ba isotopic ratio is not trivial, it is feasible,
and we intend to provide our results to compare them with future measurements,
which will provide an important test for our models.

The paper is organized as follows: in Section 2 we describe the
observational data; Section 3 describes the chemical model and the
adopted stellar yields. In Section 4 our results are
presented, and in Section 5 we summarize our conclusions.

\section{Observational data}

For the halo we employed  the same data  as presented in \citet{Cescutti13},
where we have adopted observational abundance ratios from the
literature; the data for the neutron capture elements and for the
$\alpha$-elements are those compiled by
\citet{Frebel10}\footnote{http://cdsarc.u-strasbg.fr/cgi-bin/qcat?J/AN/331/474},
labeled as halo stars\footnote{The list of authors we use from the
  collection are \citet{MCW95}, \citet{MCW98}, \citet{ WES00}, \citet{
    AOK02b}, \citet{ COW02}, \citet{ IVA03}, \citet{ Honda04},
  \citet{AOK05}, \citet{ BAR05}, \citet{ AOK06b}, \citet{ IVA06},
  \citet{ MAS06}, \citet{ PRE06}, \citet{ AOK07c}, \citet{ Franc07},
  \citet{ LAI07}, \citet{COH08}, \citet{ LAI08}, \citet{ ROE08},
  \citet{ BON09}, \citet{HAY09}}. To this sample, we added the very
recent data measured by \citet{Aoki13}.  Among the halo stars
collected, we differentiate the normal stars from the carbon-enhanced
metal-poor (CEMP) stars.  Around 20\% of stars with [Fe/H] $< -$2.0
are CEMP stars \citep{Lucatello06}.  We followed the definition given by
\citet{Masseron10}, where a CEMP star is defined as having
[C/Fe]$>$0.9.  The important distinction in the present work is made
  between CEMP-s (including CEMP-rs) and CEMP-no (including CEMP-r),
  that is, we distinguished whether a strong signature of s-process is present.

Indeed, CEMP-s (and CEMP-rs) stars most likely stem from binary mass transfer from a
previous asymptotic giant branch companion \citep{Bisterzo12,Lugaro12}, and for this
reason CEMP-s (and CEMP-rs) do not reflect the chemical evolution of the
ISM.  We  therefore opted to exclude the CEMP-s (and CEMP-rs)
stars from our figures  and included only the CEMP-no and the CEMP-r stars; in the compilation by
Frebel and in the data reported by \citet{Aoki13} there is also an
large portion of stars without carbon measurements. For these
stars we cannot establish whether they are CEMP stars or not; still,
since they represent a large portion we decided to include them in our
plots, but to distinguish them graphically from the confirmed normal
stars.

Because the data come from different authors, the methods, instruments and quality of the spectra are not
homogenous. Nevertheless, the number of metal-poor stars for which detailed abundances are available is still
impressive:  
we have
   found measurements of Ba abundances for 774 stars in the
   literature. For 459 of these, carbon abundances have also been
   determined, and 67 of the objects are classified as CEMP stars. Of
   these 67 stars, 21 are classified as CEMP-no, and one as a CEMP-r
   star.

Our sample is clearly biased toward extremely low metallicity: there
are more stars (with carbon measurement) with an [Fe/H] $<-$2.5 than
with an higher ratio, which is at odds with the metallicity distribution
function of the Galactic halo. This simply reflects the observational
strategies; typically, the most metal-poor candidates are selected a
priori using low-resolution spectra or photometry, and then these
stars are followed-up with time-consuming high-resolution
observations. Other biases can also play a role; for example, a preferential selection toward high (or low) abundance ratios for
neutron capture elements. Indeed, for certain elements (such as Eu) the
lines tend to be very weak and only upper limit detections are
available if the abundance is below  a certain threshold value
  (which is also a function of the signal-to-noise ratio of the
  spectra). We kept these biases in mind when presenting
  our results.

\section{Chemical evolution model}

The chemical evolution model presented here is the same as in
\citet{Cescutti13}.   Therefore, we describe its main
characteristics only briefly.

We considered the same chemical evolution model as adopted in
\citet{Cesc10}, which is based on the inhomogenous model developed by
\citet{Cesc08} and on the homogeneous model of \cite{Chiappini08}. 
The halo consists of many independent regions, each with
the same typical volume, and each region does not interact with the
others.  Accordingly, the dimension of the volume is expected to be large enough to
allow us to neglect the interactions between different volumes, at least
as a first approximation. For typical ISM densities, a
supernova remnant becomes indistinguishable from the ISM -- that is,
merges with the ISM -- before reaching $\sim50pc$ \citep{Thornton98}
 therefore, we decided to have a typical volume with a radius of roughly
90 pc, and the number of assumed volumes is 100 to ensure good
statistical results.  We did not use larger volumes because
we would lose the stochasticity we are looking for; in fact, larger
volumes produce more homogeneous results.

In each region, we assumed the same law for the infall of the gas with
primordial composition, following the homogeneous model by
\citet{Chiappini08}:

\begin{equation}
\frac{dGas_{in}(t)}{dt} \propto e^{-(t-t_{o})^{2}/\sigma_{o}^{2}}, 
\end{equation}
where $t_{o}$ is set to 100 Myr and $\sigma_{o}$ is 50Myr.
Similarly, the star formation rate (SFR) is defined as
\begin{equation}
SFR(t) \propto ({\rho_{gas}(t)})^{1.5},
\end{equation}
where $\rho_{gas}(t)$ is the density of the gas mass inside the  volume under consideration.
Moreover, the model takes an outflow
from the system into account:
 \begin{equation}
\frac{dGas_{out}(t)}{dt}  \propto SFR(t).
\end{equation}

Knowing the mass that is transformed into stars in a time-step
(hereafter, $M_{stars}^{new}$), we therefore assigned the mass to one
star with a random function, weighted according to the initial mass
function (IMF) of \citet{Scalo86} in the range between 0.1 and
100$M_{\odot}$.  We then extracted the mass of another star and
repeated this cycle until the total mass of newly formed stars
exceeded $M_{stars}^{new}$.  In this way, the  $M_{stars}^{new}$ is the
same, in each region at each time-step, but the total number and mass
distribution of the stars are different. We thus know the mass of each
star contained in each region, when it is born, and when it will die,
assuming the stellar lifetimes of \citet{MM89}.  At the end of its
lifetime, each star enriches the ISM with its newly produced chemical
elements and with the elements locked in that star when it was formed,
excluding the fractions of the elements that are permanently locked
in the remnant.

Yields for Fe are the same as in \citet{Franc04}.  The model considers
the production by s-process from 1.5 to 3M$\odot$ stars and SNIa
enrichment, as in \citet{Cesc06}.  

As shown in \citet{Cescutti13}, our model is able to reproduce the MDF
measured for the halo by \citet{Li10};  the results by
  \citet{Li10} are based on main-sequence turn-off stars from the data
  of the Hamburg/ESO objective-prism survey (HES)
  \citep{Christlieb08}.  This comparison shows that the timescale of
enrichment of the model is compatible with  that of the halo stars
in the solar vicinity. Moreover, our model predicts a small spread for
the $\alpha$-elements Ca and Si, compatible with the observational
data.  In addition, our model  predicts  a slight increase in
the scatter in the very metal-poor regime.  Interestingly,
inhomogeneous models, as the one shown here, do predict a few outliers
with low values of [$\alpha$/Fe] \citep[see Fig. 2 of ][]{Cescutti13}.

\subsection{Stellar yields for heavy elements}

\subsubsection{\emph{Empirical} yields for the r-process}

We considered two scenarios for the r-process.  The first is the same as
the r-process scenario assumed in \citet{Cescutti13}: the
stellar mass range contributing to the r-process is from 8 to
10~\msun. This scenario is similar to the one proposed by theoretical
models of electron-capture supernovae (EC SNe).  We call this model
\emph{EC+s}. We added +s to the names of all models to
indicate that we included the contribution by
spinstars (discussed in the next section). The yields are calculated
 with the following approach \citep[see also][]{Cesc06}: we
computed a homogeneous chemical evolution model where the yields of Ba
were chosen so as to reproduce the mean trend of [Ba/Fe] versus [Fe/H],
see Fig.~\ref{fig3} in \citet{Cescutti13}.

The second scenario follows the idea described in
\citet{Winteler12}. According to these authors, a small percentage of
massive stars end their lives as magnetorotationally driven (MRD)
supernovae.  To implement this scenario into our chemical evolution
model, we assumed that only 10\% of all the simulated massive
stars  contribute to the r-process.  This percentage is higher
than the rough  estimate of 1\% of the massive stars
mentioned in \citet{Woosley06} at solar metallicity, which is the
reference of \citet{Winteler12}, but this number is expected to
increase toward lower metallicity \citep{Woosley06}.

Similarly to the EC SNe scenario, for the MRD SNe scenario there
is no prediction of the ejected mass in each r-process event. 
On these grounds we used  the constraints computed for the
EC SNe scenario on the basis of observational data; we present two
cases for the r-process yields of the MRD SNe scenario:

\begin{itemize}
\item \emph{MRD+s~A}: in this case we made the simplest assumption
  that 10\% of all the stars with masses between 10~\msun~and
  80~\msun~  generate an r-process that produces a fixed
    mass. We determined the ejected mass of Ba for each event in such
  a way that \emph{MRD+s~A} produces the same amount of Ba as the
  \emph{EC+s} model in a stellar generation.  The  resulting yield of r-process
  Ba is 8.0 $\cdot 10^{-6}$\msun; the other chemical
  elements were simply scaled using the solar system r-process
  contribution as determined by \citet{SSC04}.

\item \emph{MRD+s~B}: in this case we took into account the
  possibility that the amount of mass ejected as r-process varies.
 Since the variation is unknown, we assumed this range:
  the minimum is 1\% of the fixed value
  in the previous model and the maximum is twice the same value.
  Since the total production should be conserved, the ejected mass for
  n- star (r-process producer) in the model \emph{MRD+s~B} can be
  described by the following equation:
\begin{equation}
M_{Ba}^{MRD+s~B}(n) = M_{Ba}^{MRD+s~A}~(0.01+1.98\cdot Rand(n)), 
\end{equation}
 where Rand(n) is a  uniform random distribution in the range [0,1].

\end{itemize} 
The models \emph{EC+s}, \emph{MRD+s~A} and \emph{MRD+s~B} are
summarized in Table \ref{table1}.

\begin{table*}
\begin{center}
\caption{Nucleosynthesis prescriptions for the three cases analyzed for strontium.}

\label{table1}
\begin{tabular}{|c|c|c|c|}
 \hline
  Model name&rows in Fig. \ref{fig1}    &  r-process  site  &  r-process ejecta \\  
\hline
  EC+s  & first    &  all the stars between 8-10~\msun  &  mass dependent \citep[see Fig.4][]{Cescutti13}\\
\hline
  MRD+s~A & second   & 10\% of stars between 10-80~\msun & a fixed value \\
\hline
  MRD+s~B & third     &  10\% of stars between 10-80~\msun & 1\% and 199\% of the value of MRD+s  A \\
\hline 
 \end{tabular}
\end{center}

\end{table*}

The difference between the r-process ratio observed in
metal-poor r-process-rich stars \citep[see][]{Sneden08} and the solar
system r-process contribution as determined by \citet{SSC04} 
seriously affects on the Y ratio, which is 0.5 dex lower in the observational
case.   Therefore, we computed other two models with the
only difference  that we assumed the r-process ratio observed in
metal-poor r-process-rich stars \citep[see][]{Sneden08}, instead of 
the solar system r-process contribution determined by \citet{SSC04} as
in the previous scenarios.  The models are denoted
\begin{itemize}
\item \emph{EC+s~2}: the model based on the EC+s  scenario with
  the observed r-process ratios in r-process-rich stars,
\item \emph{MRD+s~B2}: the model based on the \emph{MRD+s~B} scenario
  with the observed r-process ratios in r-process-rich stars.
\end{itemize} 
These two models and the comparison model \emph{EC+s} are summarized
in Table \ref{table2}.

\begin{table*}
\begin{center}
\caption{Nucleosynthesis prescriptions for the three cases analyzed for yttrium.}

\label{table2}
\begin{tabular}{|c|c|c|c|}
 \hline
  Model name& rows in Fig. \ref{fig3}    &  r-process  site  &  ratio Y/Ba \\
\hline
  EC+s  & first   &  all the stars between 8-10~\msun &  solar residual \citep{SSC04}\\
\hline
  EC+s~2 & second    &  all the stars between 8-10~\msun &  r-process-rich halo stars \citep{Sneden08}\\
\hline
  MRD+s~B2 & third     &  10\% of stars between 10-80~\msun & r-process-rich halo stars \citep{Sneden08}\\
\hline

 \end{tabular}
\end{center}

\end{table*}

We do not show these models for the Sr, since the difference in this case
would be only of $\sim$0.1 dex, that is, lower than the typical error of 
the observational data.

\label{sec:yields}
\subsubsection{Contribution of \emph{spinstars}}

We assumed for all our models the same contribution by s-process as in the
\emph{fs-}model of \citet{Cescutti13}. However, we show here results for
yttrium which was not treated in the previous paper, therefore we recall that
for the yields at Z=$10^{-5}$ \footnote{[Fe/H]$\simeq -$3.5, with small
variations due to the stochasticity of the models.} we considered the
stellar yields obtained by \citet{Frisch12} after decreasing the reaction rate for
$^{17}O(\alpha,\gamma)$ from \citet{CF88} by a
factor of 10, which enhances the s-process  production. 
Unfortunately, we have results with this reaction rate only
for a single mass (25~\msun) at Z=$10^{-5}$, and we used the
scaling factor obtained for the whole range of masses
\citep[see][]{Cescutti13}. Indeed, there are no
nucleosynthesis calculations for spinstars currently carried out with a reduced
value of the $^{17}O(\alpha,\gamma)$ rate for a metallicity higher  than
Z=$10^{-5}$, and we adopted those computed with the standard
value given by \citet{CF88}. We need to keep this caveat in mind when
interpreting our theoretical predictions for the intermediate
metallicity range. Finally, we note that in the spinstars framework,
Eu is produced in negligible amounts.

\section{Results}

\subsection{Stellar distribution for strontium and barium}

\begin{figure*}[ht!]
\begin{minipage}{185mm}
\begin{center}

\includegraphics[width=185mm]{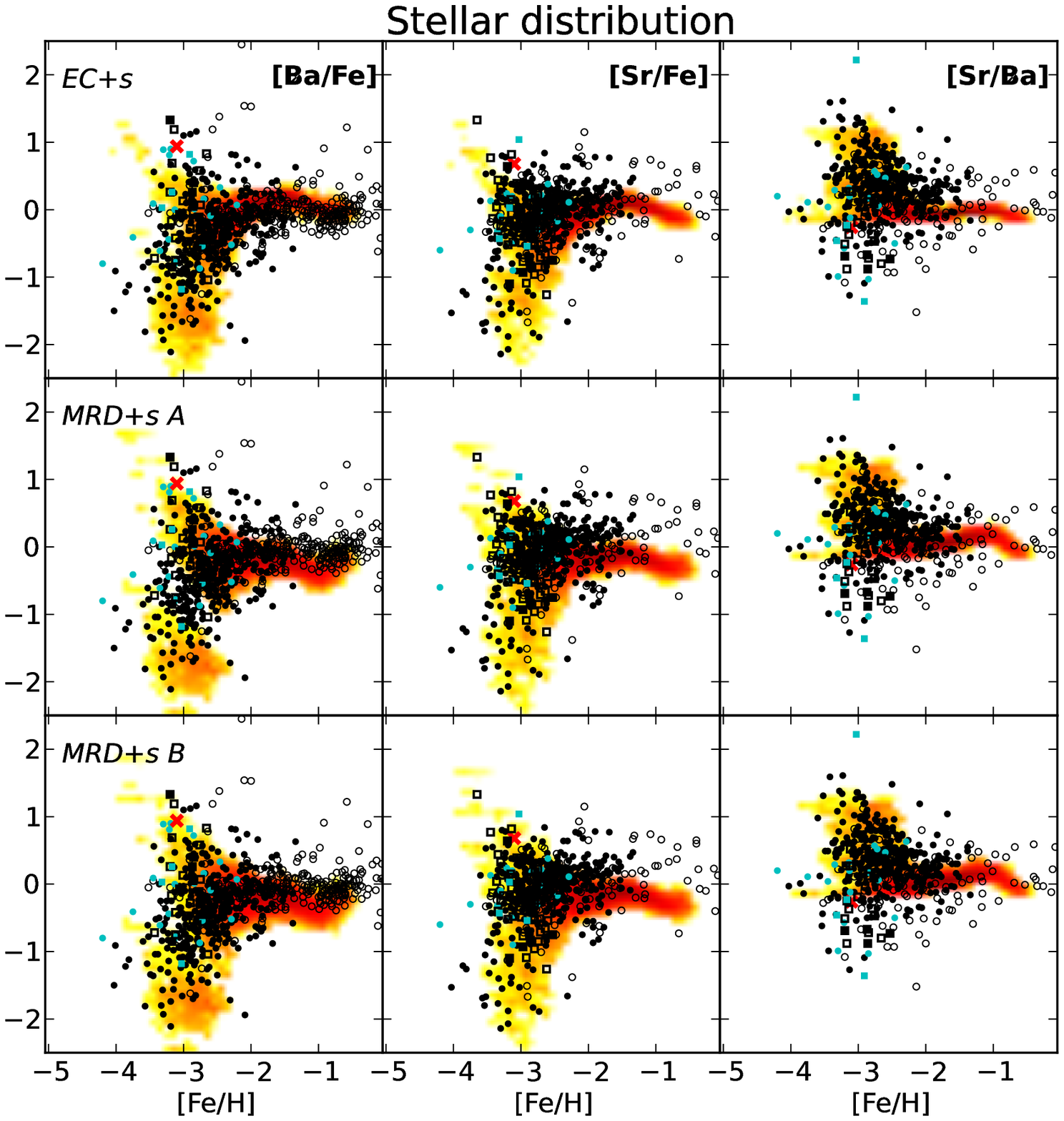}
\includegraphics[width=160mm]{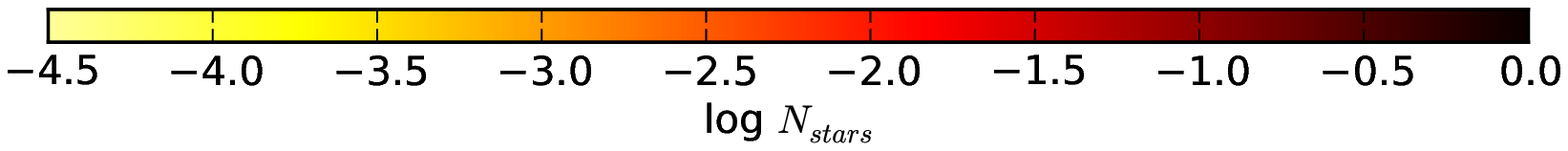}

\caption{From the left [Ba/Fe], [Sr/Fe] and [Sr/Ba] vs [Fe/H] in the
  halo; the density plot is the distribution of simulated long-living
  stars for our models, see bar over the main figure for the color
  scale; superimposed, we show the abundances ratios for halo stars
  \citep[data from][]{Frebel10,Aoki13}. The symbols for the
  \citet{Frebel10} data are black dots for normal stars, cyan dots for
  CEMP-no, a red x marker for the CEMP-r star, and black open circle
  for stars without carbon measurement; for the \citet{Aoki13} data we
  adopt the same symbols, but instead of dots we used squares.}\label{fig1}

\end{center}
\end{minipage}
\end{figure*}

Fig.\ref{fig1} summarizes our results for the stellar
distribution of strontium and barium in the Milky Way halo. In the first
  row we reproduce the results obtained for the \emph{fs-}model of
  \citet{Cescutti13} (see their Fig. 5), in which the r-process is
  assumed to be produced by stars in a narrow mass range (the EC
  scenario).  In the second row, we present the \emph{MRD+s~A} model,
  which assumes an MRD scenario for r-process. In particular, this
  model is computed with the assumption that 10\% of all the
  massive stars contribute to the r-process enrichment. The s-process
production remains the same as in the model above.

Overall, the comparison with the results of the \emph{EC+s} model does
not reveal  significant differences. One main feature is a gap in the stellar distribution in the [Ba/Fe]
vs. [Fe/H] diagram.  This gap is due to the significant difference we
assumed between the production of Ba in spinstars and in the
r-process events; this gap is absent from the [Sr/Fe] plot for which the two
stellar yields are of the same order of magnitude.  The fact that a
gap is not seen in the data suggests that our assumption that the
amount of  mass produced by the r-process is
independent of the stellar mass or any other parameter (for
instance, magnetic field and rotation) is too simplistic.  

In model \emph{MRD+s~B} we relaxed the assumption of a constant yield
for the r-process, and allowed  a variation in the total amount of
r-process  production (see third row in Fig.\ref{fig1}).  In this case,
the gap in the [Ba/Fe] distribution disappears. In addition, model
\emph{MRD+s~B} predicts a slightly more pronounced spread in the
[Ba/Fe] distribution at intermediate metallicities
($-$2$<$[Fe/H]$<-$1).

We now focus on the comparison between \emph{MRD+s~B} and the
\emph{EC+s} models (first and third rows in Fig.~\ref{fig1}). The two
models give very similar results. In other words, the observational
constraints  displayed in this figure cannot distinguish between these
two scenarios. In both scenarios the time delay in the
r-process production is short: the first star (r-process producer) in the
MRD scenario explodes after 3-4Myr
(lifetime of a 80~\msun~star), whereas in the case of the EC scenario this
occurs after 20~Myr (lifetime of a 10~\msun~star).  These timescales
coupled with the star formation of the Galactic halo do not produce
any appreciable difference. We expect, however, that in a system with
a more intense star formation history the small differences in the
time delay of these two cases can lead to differences in the
predicted distributions for the heavy element abundance ratios that
might  in principle  be testable through observations. We will explore this
possibility when studying the halo versus bulge chemical enrichment,
in a forthcoming paper.

In Fig.\ref{fig1}, some of the observational data present a
  [Sr/Ba] ratio that is lower than our model predictions. The reason
  for this is that our models only include the r-process contribution
  (as given by the solar system pattern) and the s-process
  contribution by spinstars. The spinstars contribution \citep[with
  the current stellar yields taken from][at the end of the pre-SN
  phase]{Frisch12} produces [Sr/Ba] ratios higher than the solar
  system value. The situation can be more complex, however, as shown
  in \citet{Chiappini11}, where stellar yields were computed at two
  different phases of He-burning (see their figure 2), showing that
  there might be situations where spinstars produce a negative [Sr/Ba]
  ratio. More detailed stellar yields are needed to investigate this
  point in more detail.  In addition, the scaled r-process
  contribution adopted here is a simplified approach (see discussion
  on the [Y/Ba] ratio).  Finally, some objects might still have a
  small contribution from AGB mass-transfer, which would again
  decrease the [Sr/Ba] ratios. The AGB mass-transfer contribution is
  not taken into account in our models. However, only very few data
  points show negative [Sr/Ba] ratios (see also Fig.~\ref{fig5}) .  It
  is encouraging that with only two nucleosynthesis sites for Sr and
  Ba, our results agree so well with the metal-poor data. 

\subsection{Isotopic distribution of Ba}

\begin{figure*}[ht!]
\begin{minipage}{185mm}
\begin{center}

\includegraphics[width=185mm]{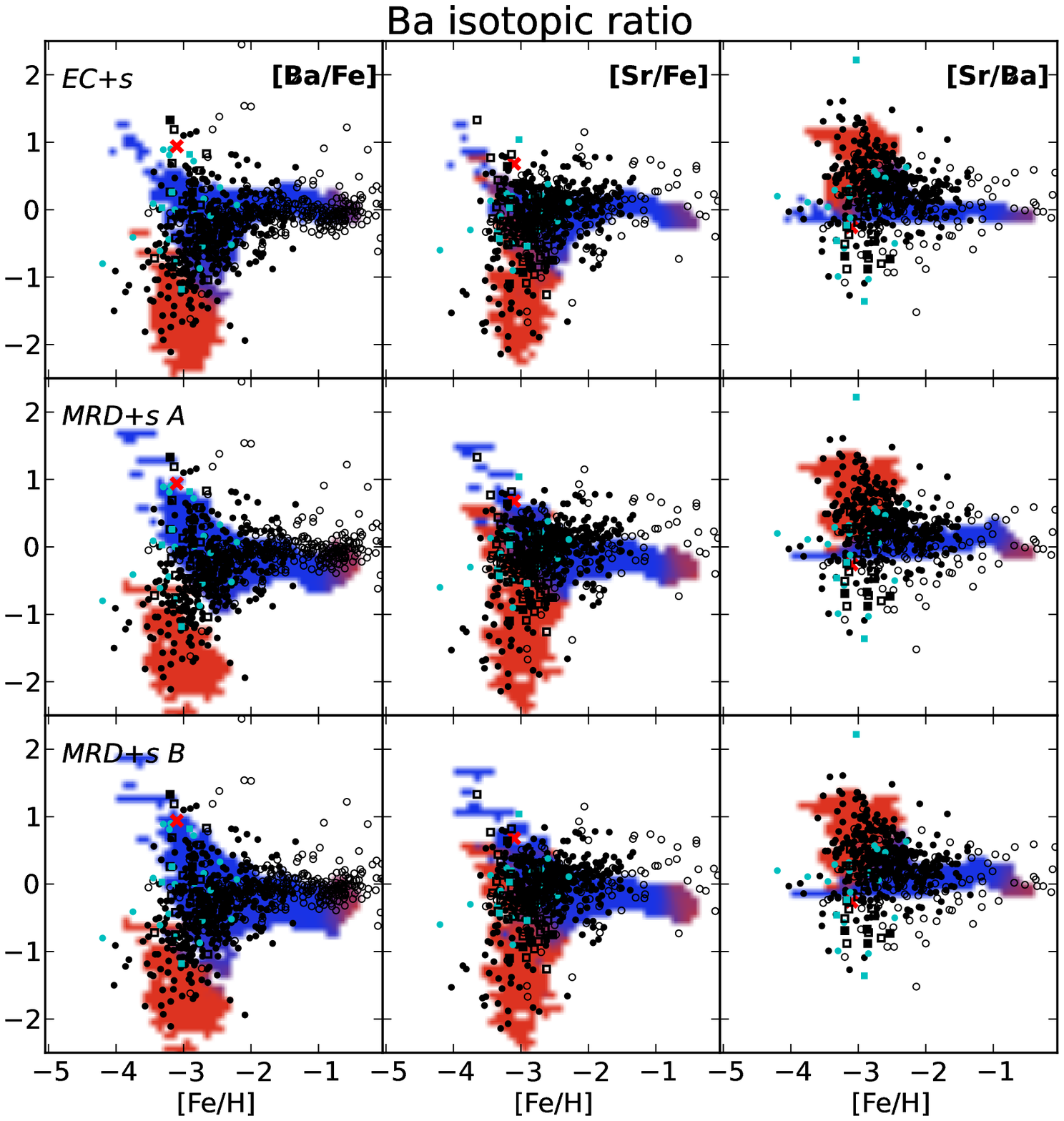}

\includegraphics[width=160mm]{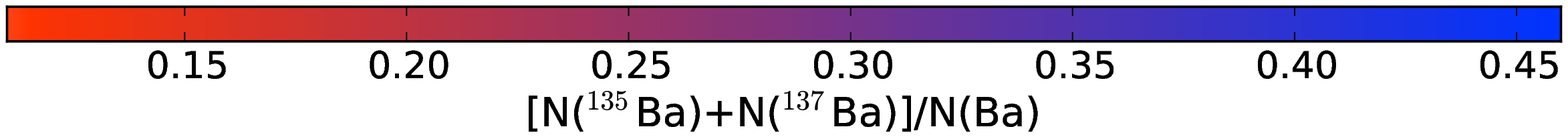}

\caption{From the left [Ba/Fe], [Sr/Fe] and [Sr/Ba] vs [Fe/H] in the
  halo.  The density plot is the distribution of the isotopic ratio of
  Ba according to the models, see the bar below the main
  figure. The data are the same as in Fig.~\ref{fig1}.
}\label{fig2}
\end{center}
\end{minipage}
\end{figure*}

In Fig.~\ref{fig2}, we present our model predictions for the isotopic
ratio of Ba in Galactic halo stars for the same models as presented in
Fig.~\ref{fig1}. This is the first time these results have been
computed, and even though the measurement of the Ba isotopic ratio is
challenging, it provides an important method to test our model
predictions.  We are also interested in investigating whether the Ba
isotopic ratio might  help in distinguishing the two scenarios for
r-process production studied here.

The color code in Fig.~\ref{fig2} indicates the fraction of odd
isotopes of Ba
\footnote{$\frac{N(^{135}Ba)+N(^{137}Ba)}{N(Ba_{tot})}$} in the
simulated stars.  The ratio of odd isotopes can be used as an
indicator for the presence of the r-process production (blue
color) or the s-process production (red) in our simulated stars.
Solid nucleosynthesis results indeed indicate that the s-process
preferentially produces even isotopes, thus generating a low odd
fraction of Ba isotopes \citep[0.11 according to][]{Arla99};
conversely the r-process has no preferential production between odd
and even isotopes, thus generating an odd fraction of about 0.5
\citep[0.46 according to][]{Arla99}.

In the three first plots in the first row, we show the results for the
\emph{EC+s } r-process site.  As expected, there is a strong r-process
signature in the stars with a high [Ba/Fe] (or [Sr/Fe]), whereas a
strong s-process signature marks the stars with low [Ba/Fe] ([Sr/Fe]).
The two populations are present and rather distinct up to [Fe/H]$\sim
-$2.5, above which the r-process signature becomes dominant. This is
due to the larger mass produced by the r-process compared with
s-process by spinstars. As soon as an r-process event enriches the
simulated volume, the isotopic ratio tends to be r-process-like.  At
the very end of the simulation, the production of s-process by
low-intermediate mass stars moves the isotopic ratio toward an
intermediate situation with an odd fraction of $\sim$0.3.  For the
[Sr/Fe] case, where the yields are of the same order, this effect is
less pronounced.  Indeed, the larger production of Sr by spinstars
(compared to Ba) explains why there are extremely metal-poor stars
with a strong s-process signature at [Sr/Fe]$>$0 in the [Sr/Fe]
diagram (in contrast to what is seen for the [Ba/Fe] ratio).  In this
area an intermediate isotopic ratio is displayed as well; this is due
to stars with s-process and r-process signatures that share the same
locus on the graph and not to a real mixture in each star.

The situation is different for the [Sr/Ba] plot. In this  case
the stars dominated by r-process nucleosynthesis lie on the r-process
ratio assumed for our empirical yields, whereas the stars above 
  it  exhibit an enrichment by spinstars and are thus s-process-rich
(i.e., with a low fraction of odd Ba isotopes).  This plot 
  clearly shows the impact of spinstars on promoting the [Sr/Ba] scatter
  observed in the observational data.

  The second- and third- row models \emph{MRD+s~A} and \emph{MRD+s~B} - show
  very similar results in this figure as well.  Therefore, we 
  underline the differences between the \emph{MRD+s~B} and the \emph{EC+s}
  models. The different r-processes considered lead to only small
  differences in the fraction of s-process-dominated loci in all the
  plots. This means that it is not possible to use the Ba isotopic ratio to
  distinguish between the two possibilities.

  In summary, the main result is that in the spinstars framework we
  expect a large portion of extremely metal-poor stars to show a
  clear s-process signature. The latter tends to be quickly erased by
  the larger injection of r-process material as soon as their sites
  become fully operative (in scales on the order of 10~Myrs). This is
  a robust prediction regardless of the nature of the site for the
  r-process production.

\subsection{Stellar distribution for yttrium}

\begin{figure*}[ht!]
\begin{minipage}{185mm}
\begin{center}
\includegraphics[width=185mm]{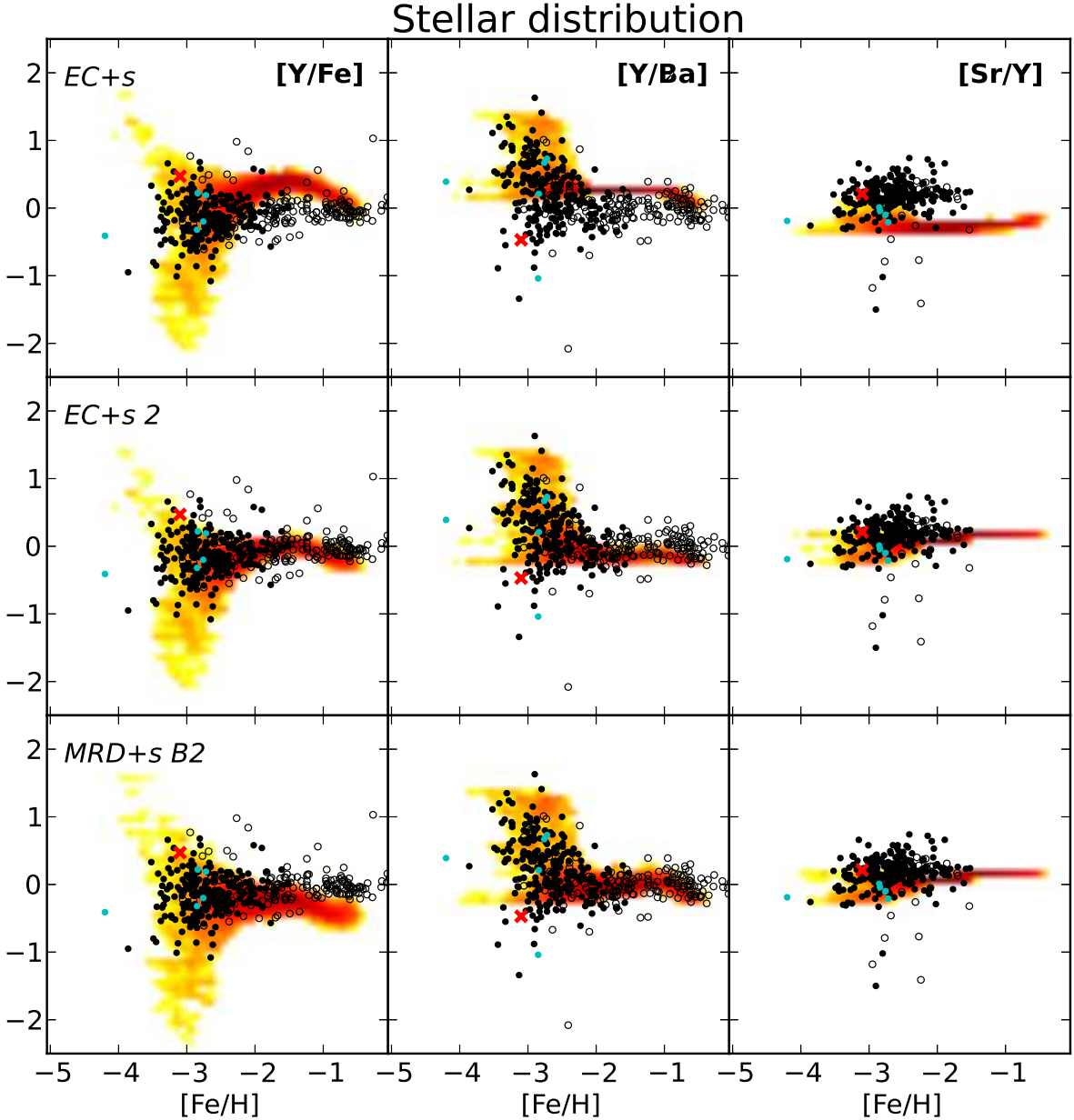}
\caption {From the left [Ba/Fe], [Y/Fe] and [Y/Ba] vs [Fe/H in the halo; the
  density plot is the distribution of simulated long-living stars for
  our halo models, see bar below Fig.~\ref{fig1} for the color scale; the data
are the same as in Fig.~\ref{fig1}.}\label{fig3}
\end{center}
\end{minipage}
\end{figure*}

In Fig.~\ref{fig3} we show our results for another light neutron
capture element for which observational data are present in
literature: Y.  The comparison with a new element is important because
we can further test our theory by verifying whether our results for
the new element agree with the observational data using the same
nucleosynthesis hypothesis.

The comparison of the \emph{EC+s} model with the
observational data for [Y/Fe] and [Y/Ba] is not satisfactory. In this
case we adopted the same ingredients for s- and r-process
as for the Sr \emph{EC+s} model. We note that for [Fe/H]$>-$2,
where the model result distribution is  no longer affected by a
strong spread, the model results appear to be roughly 0.5 dex above
the observational data; this is  particularly visible in the [Y/Ba]
ratio.  In this regime, as we have learned from the isotopic
distribution of Ba, the dominant contributor to the chemical evolution is the r-process.
On the other hand, we recall that we assumed the
solar system r-process pattern of \citet{SSC04} for the ratios between Ba
(which is fixed empirically), Y and Sr.

To estimate the solar system r-process contribution \citet{SSC04}
removed the s-process contribution produced by low-intermediate
mass stars from the solar abundances. In addition to possible
uncertainties in accounting for the s-process contribution by low-mass
stars (and the integration to find the solar abundances), other
contributors such as spinstars, might be hidden in the so-called
r-process residual when this method is adopted.  An alternative way to
infer the r-process pattern is to use the observed ratio between Sr
and Ba in r-process-rich stars \citep{Sneden08}. This method is
affected by observational errors, which are kept low by averaging five
of the known r-process-rich stars. With the observational fraction the
Y/Ba is subject to a displacement of -0.5 dex.  The effect of this
different method is shown in the second row of Fig.~\ref{fig3}, which
displays the results obtained with the \emph{EC+s~2} model in which we
assumed the observed r-process signature (see also Table 2).  For the
\emph{EC+s~2} model our results agree very well with the observational
data.  Finally, for Y we predict a slightly smaller spread than for
[Sr/Fe] (and [Ba/Fe]).  The same is shown by the observational
data. This result arises naturally from the higher amount of Y
theoretically predicted by s-process in spinstars compared with Sr (and
Ba).

In the third row we show the results obtained with the model
\emph{MRD+s~B2} (we do not show the results for model \emph{MRD+s~B}
because it suffers from the same offset we discussed when presenting the
\emph{EC+s} model results).  Our conclusions here do not differ from
those indicated when we described the results for Sr. Note also that for yttrium
we cannot easily distinguish between the MRD+s and the EC+s scenarios,
and again the MRD+s scenario presents a slightly larger spread at
[Fe/H]$>-$2.

\subsection{Stellar distribution for europium}

\begin{figure*}[ht!]
\begin{minipage}{185mm}
\begin{center}
\includegraphics[width=185mm]{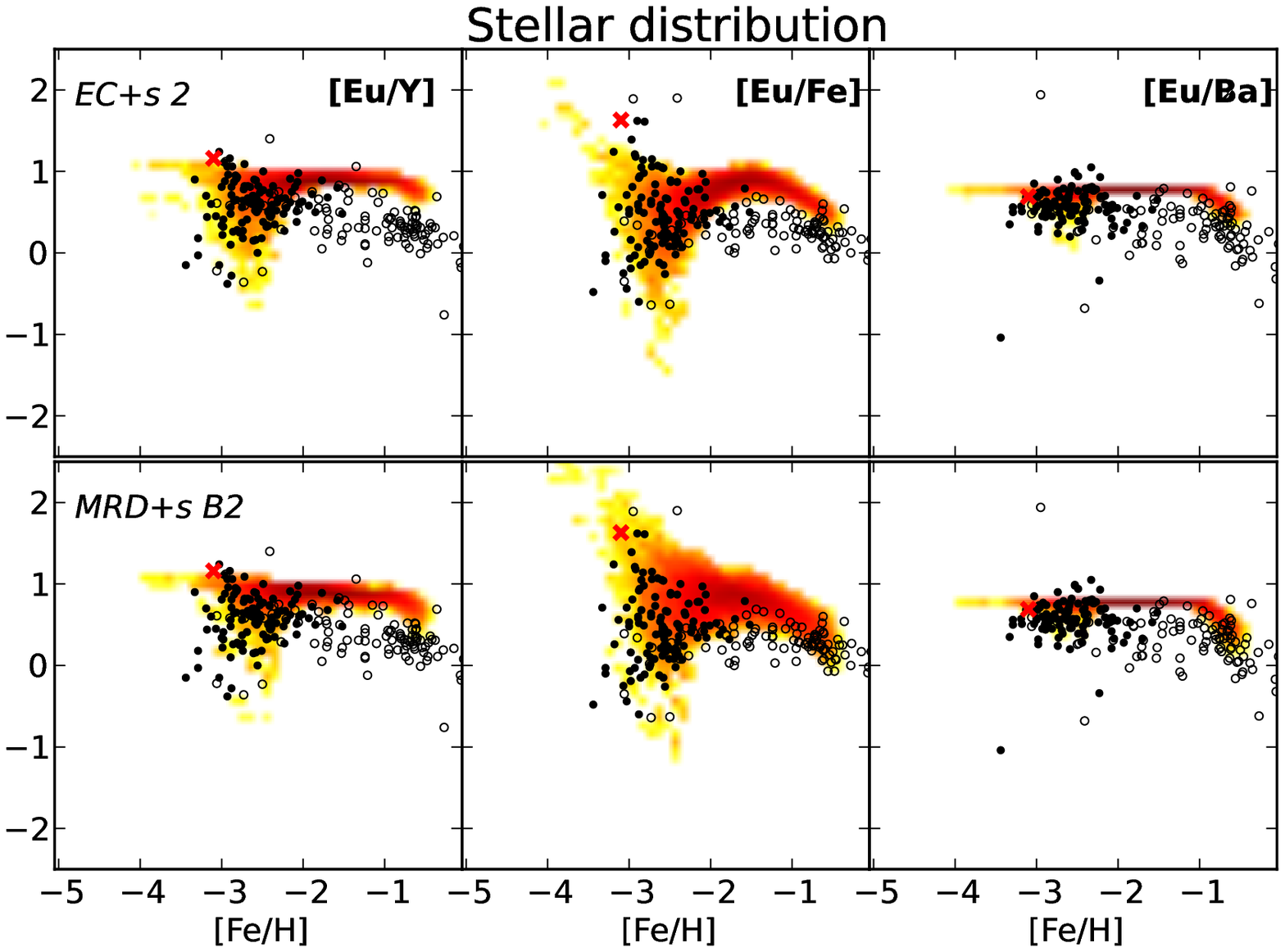}
\caption {From the left [Eu/Y], [Eu/Fe] and [Eu/Ba] vs [Fe/H] in the
  halo; the density plot is the distribution of simulated long-living
  stars for our halo models, see bar below Fig.~\ref{fig1} for the
  color scale; the data are the same as Fig.~\ref{fig1}.}\label{fig4}
\end{center}
\end{minipage}
\end{figure*}

In Fig.~\ref{fig4} we present the results for the neutron capture
elements Eu compared with Y, Fe, and Ba.  The element Eu is a heavy
neutron capture element (its stable isotopes have atomic masses 151
and 153) and peculiar in that it is produced almost entirely by the
r-process (typically less than 5\% is believed to be produced through
the s-process at solar metallicity). The Eu lines are weak, and in the
literature we found only upper limits for the stars at [Fe/H]
$<-$3.

Although Eu is not produced by spinstars according to the current models,
this element can be important to distinguish between the r-process
scenarios because it is   uncontaminated by other production sites. Moreover, it
can be useful as a reference element, instead of iron.

For the r-process production, the [Eu/Fe] ratio
shows the largest discrepancy between the two
scenarios (see the central column of Fig.~\ref{fig4}).  To have enough
spread at extremely low metallicity, we need to have  a large
  r-process production (and thus Eu) in each event of EC SN (plot
above). At intermediate metallicity this produces a  too high ratio of
[Eu/Fe].  The MRD+s scenario appears to agree better
with the [Eu/Fe] ratio of halo stars.

The plots in the left column in Fig.~\ref{fig4} display the ratio
[Eu/Y] for the \emph{EC+s~2} and \emph{MRD+s~B2} models. This ratio is
interesting because Y is the element most
affected by spinstars  s-process production among the investigated
elements, whereas we do not consider any production of Eu by
spinstars. On the other hand, this ratio does not help to highlight
different features between the different r-process scenarios because we assumed 
the same r-process pattern in both cases.  The spread that we
predict for the two models agrees very well with the spread
observed in halo stars. The trend toward higher metallicity is
slightly different between the two models: the MRD+s scenario tends to
keep a larger spread in the [Eu/Y] ratio even at metallicities above
$-$2.5. Both models show a smooth decrease of the [Eu/Y] ratio toward
solar (at around metallicities [Fe/H] = $-$1) because of the s-process
production in intermediate-mass stars. Although the data also show a
decreasing [Eu/Y] ratio toward higher metallicities, the observations
are systematically lower than the model predictions. How important  this
discrepancy is, is difficult to access because for this
metallicity range, there is no information on the carbon abundance in
these stars (as illustrated by the open symbols). Therefore, we cannot
exclude that the abundances observed in some of these stars are
contaminated by a binary star; in this case, material enriched in Y
(and other elements predominantly produced in the s-process)
coming from the binary star can change the original ratio of the
chemical elements so that the observed ratio of [Eu/Y] is expected to
be lower than the predictions from our chemical evolution models.

The same lack of carbon measurements also afflicts the plots for
[Eu/Ba] (last column of Fig.~\ref{fig4}). To a first approximation, the
same as discussed above for Y applies for Ba; though for Ba the spread
in the [Eu/Ba] for the models is relatively small, in agreement with
the measured halo stars, again both models appear to predict a too
high ratio for [Eu/Ba] at intermediate metallicities
($-$2$<$[Fe/H]$<-$1). At these metallicities the MRD+s scenario
presents a slightly higher spread. The general trend of both models  does not satisfactorily reproduce
the observations, suggesting that a higher
production of Ba by spinstars probably also takes place at these
intermediate metallicities  to improve the match between the
predicted and observed [Eu/Ba] ratios.
We recall that we adopted the spinstar nucleosynthesis
prescriptions computed with a decreased $^{17}O(\alpha,\gamma)$
reaction rate only for Z$<10^{-5}$; since the decrease of the reaction
rate would promote the s-process production, we can expect to produce
more Ba and to alleviate this problem. Unfortunately, we can only speculate
upon this option because  for higher
metallicities we do not have nucleosynthesis results with the
decreased reaction rate at the moment.

\subsection{Distribution functions}

\begin{figure*}[ht!]
\begin{minipage}{185mm}
\begin{center}
\includegraphics[width=185mm]{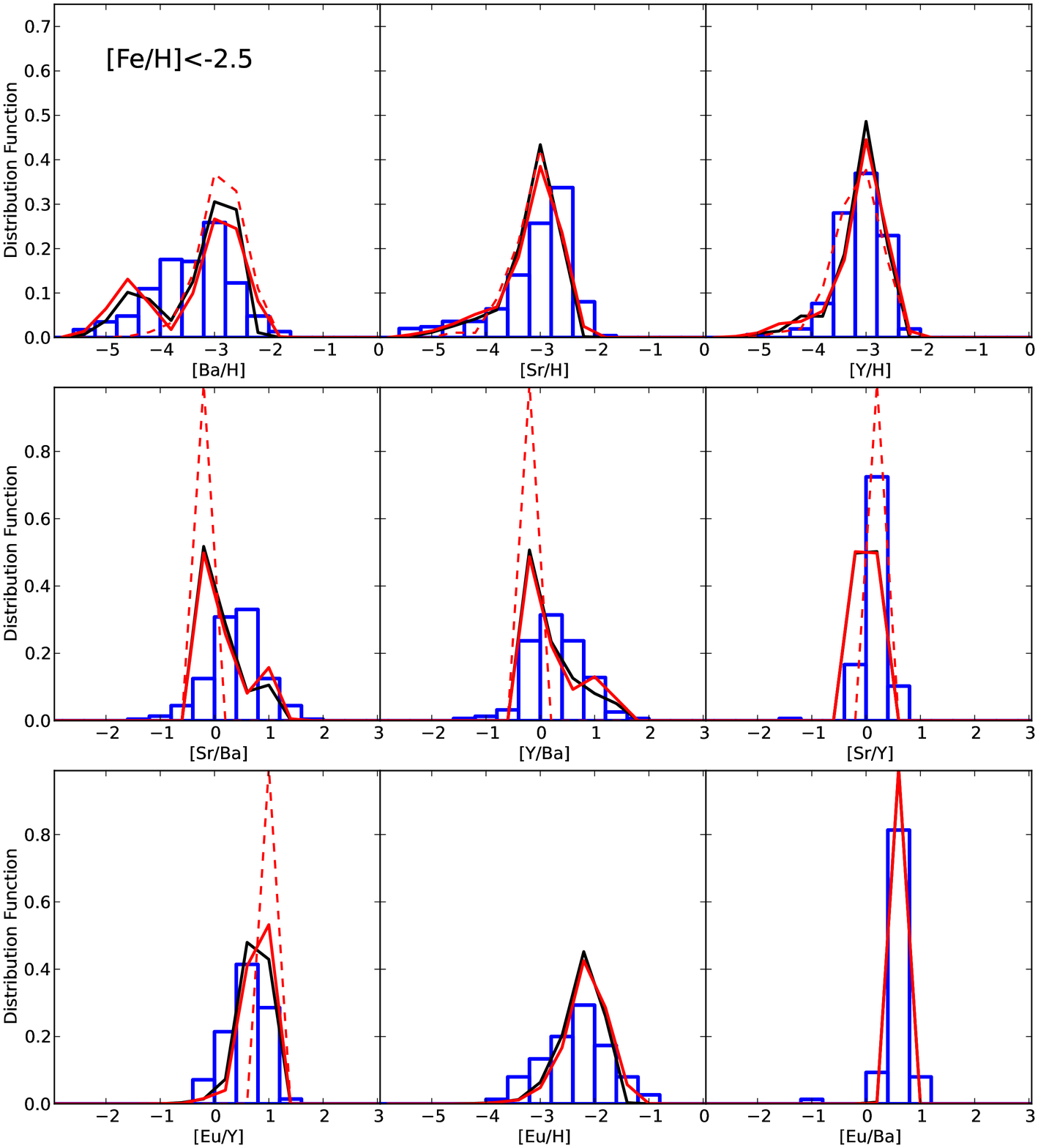}
\caption {Distribution functions of observed (blue histograms) and
  simulated stars in the range [Fe/H]$<-$2.5.  We consider nine
  different chemical abundance ratios.  Three different models are
  plotted: EC+s (black line), MRD+s~B2 (red line) and MRD-s~B2
  (without contribution of spinstars, red dashed line).  }\label{fig5}
\end{center}
\end{minipage}
\end{figure*}

\begin{figure*}[ht!]
\begin{minipage}{185mm}
\begin{center}
\includegraphics[width=185mm]{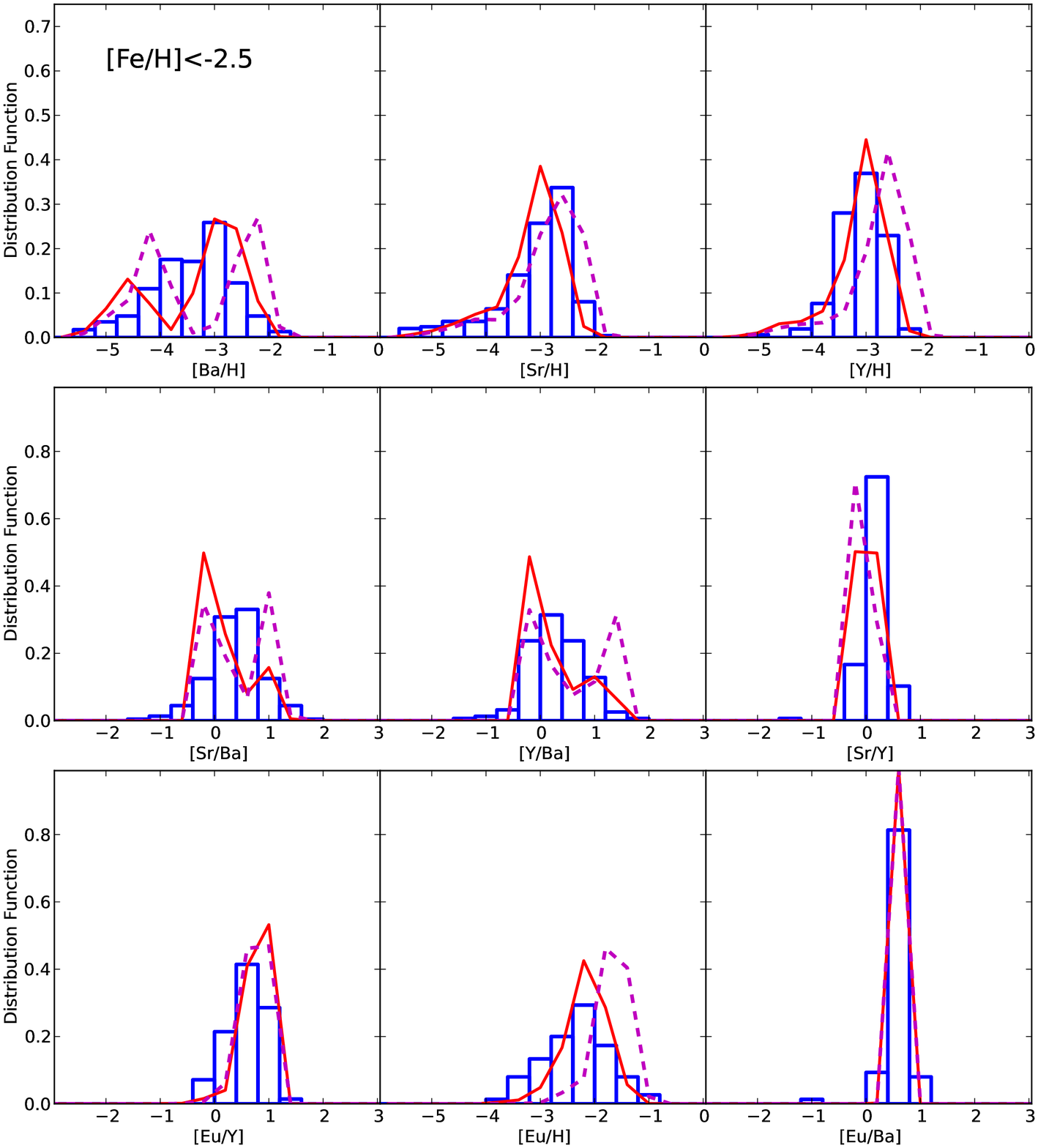}
\caption {Distribution functions of observed (blue histograms) and
  simulated stars in the range [Fe/H]$<-$2.5.  We consider nine
  different chemical abundance ratios.  In this case we plot the
  MRD+s~B2 (red solid line) and a comparison model in which we assume
  a rate of 5\% among massive stars for MRD  SNe (magenta dashed
  line).}\label{fig6}
\end{center}
\end{minipage}
\end{figure*}

The comparison of our chemical evolution model predictions (computed
with different nucleosynthesis prescriptions) with the observed
abundances in metal-poor stars (Figs.~\ref{fig1}, \ref{fig2},
\ref{fig3}, and \ref{fig4}) provides not only very useful constraints 
to the chemical enrichment of the Galactic halo, but also to the
nucleosynthetic process itself. Here we can go a step further. Tighter
constraints can be obtained by comparing the predicted and observed
distribution functions of the number of stars with a certain value of
the different studied abundance ratios. In the present section we
carry out, for the first time, to the best of our knowledge, such a
comparison \citep[except for the work of][where a similar
attempt was carried out for CNO elements]{Cesc10}. 

Thanks to the efforts in recent years of many observers, 
the number of stars measured at this extremely low metallicity regime
has increased considerably. It is now possible to obtain the
distribution (in number) of observed stars as a function of different
key chemical species with a significant statistic and compare it with
model results. However, one has to be aware of the strong biases
contained in the halo star samples (where detailed abundances are
usually obtained for the confirmed most metal-poor stars with 8-10m
class telescopes). Hence we carried out this comparison only for
metallicities [Fe/H] $< -$2.5 where the biases are certainly less
important. In addition, the selection of this metallicity range meets
our goal, which is to extract the information contained in the scatter
(or lack thereof) in the abundance ratios. The
observed spread becomes more important below [Fe/H] $\sim -$2.5. 

In Figs.~\ref{fig5} and \ref{fig6} we present nine abundance ratios
for the four neutron capture elements we studied Ba, Sr, Y and Eu: the
four elements compared with H and all the possible combination among
them (except [Eu/Sr]).  In Fig.~\ref{fig5} we present the results
obtained with three different nucleosynthesis models: the
\emph{EC+s~2}, the \emph{MRD+s~B2} (the best two models for the two
possible scenarios), and the \emph{MRD-s~B2} model (which is the
\emph{MRD+s~B2}, but without the contribution of spinstars). The
latter model then shows the predictions for the case where only the
r-process is at play. 

Before discussing the comparison with the observational data, we
note that the \emph{EC+s~2} and the \emph{MRD+s~B2} model are quite
similar, and in a distribution with a bin of 0.4 dex, there are 
tiny differences between them for [Fe/H]$<-$2.5 (note that the choice
of the bin size was made because we needed to have enough measurements
and hence statistically meaningful results per bin).  Furthermore, the
observational uncertainties can be on this order for the n-capture
elements considered in the present work.

We first focus on the absolute abundances (with respect to hydrogen)
of Ba, Sr, and Y (these are shown in the first row of
Fig.~\ref{fig5}).  The two models \emph{MRD+s~B2} and \emph{EC+s~2}
appear to agree well with the observations: our predictions for [Ba/H]
and [Y/H] match the peaks of the distributions remarkably well,
whereas for Sr there is a small offset. We remark that the peaks are
determined by the r-process contribution, and only the Ba yields for
the r-process have been empirically determined, whereas those for Sr
and Y were simply scaled using the observation of r-process-rich stars
for these models. The fact that the peaks are set by the r-process is
also illustrated by the comparison with the model MRD-s (without the
s-process contribution by spinstars): it is clear that the peaks do
not move when the contribution from spinstars is switched off.  It is
also clear that after the spinstar contribution is turned off, the
predicted distribution for the [Ba/H] becomes worse because the model
now severely underestimates the number of stars with very low [Ba/H]
ratios (the same occurs for Sr, although in this case the effect is
minor). For Y the peak position is also changed slightly. Finally, for
[Ba/H] we also note that the models with contributions from spinstars
predict a double-peaked distribution, which is not present in the
data.

In the second row of Fig.~\ref{fig5}, the plots for the [Sr/Ba] and
[Y/Ba] ratios all share similar features.  The models are able to
match the dispersion observed for these ratios thanks to the
contribution from spinstars; this result was already seen in our
previous figures. From these figures the role of spinstars in
explaining the spread for these chemical ratios is clear. It is also
clear that the r-process production by itself is unable to explain the
observed spread.  We also note that the shapes of the model
predictions do not match the observational data perfectly. In
particular for [Sr/Ba] and [Y/Ba] our models predict a peak at the
r-process ratio with a more extended tail toward high [Sr/Ba] or
[Y/Ba] ratios, whereas the data have a peak at a relatively higher
ratio with more symmetric tails.  For [Sr/Y] observations confirm the
model predictions that the scatter should be small because in this case
both elements are produced in similar ratios by r-process and
s-process.

Finally, in the third row Fig.~\ref{fig5} we present the results for
abundance ratios involving europium. It is interesting that here the
predicted and observed [Eu/Y] ratios match very well. For this ratio
the (predicted and observed) spread is smaller than that obtained for
[Sr/Ba] and [Y/Ba]. However, for [Eu/Y] there is still the tendency to
have too many modeled stars closer to the r-process signature (which
is shown by the peak of the MRD+s model without spinstars). For
[Eu/H] we are also able to match the peak and the spread; for this
element models with and without spinstars show the same results because
spinstars do not contribute Eu. It is interesting to note that in this
plot the spread for [Eu/H] is less pronounced than for [Ba/H]. For
[Eu/Ba] the spread expected by the model is smaller than the
dimensions of the assumed bin because of the dominant role of the
r-process production compared with s-process contribution by
spinstars. The lack of dispersion in the [Eu/Ba] ratio is also
confirmed by the data.  More measurements for Eu are clearly needed.

The purpose of Fig.~\ref{fig6} is to illustrate the potential of
  our approach for constraining the site of the r-process.  In this
figure we show our \emph{MRD+s~B2} model compared with a similar model
in which we decreased the probability of having a r-process event
from 10\% to 5\%.  In the latter model, to conserve the
average production of the r-process by a stellar generation, we have
increased the amount of r-process material ejected by each event by a
factor of two. This change does not affect  considerably the
comparison of the model results and data in a stellar density plot
similar to the one shown in Fig.~\ref{fig1}. This is illustrated in
the appendix in Fig.~\ref{fig7}.  In the diagrams shown in
Fig.~\ref{fig6}, the difference between the two models can be easily
appreciated.  A lower probability for the r-process contribution
implies that the distribution of stars will be more influenced by the
contribution  from spinstars. This is the case in all the diagrams of
Fig.~\ref{fig6}. From this figure the danger of
using biased samples also becomes clear. Indeed, any biases toward, for instance,
r-process-rich stars can provoke a deformation of the observed
distributions and erroneously favor a given scenario or different
rates.  Accounting for the observational biases will be mandatory in
the future of very metal-poor research if one wishes to use their
precious information to better constrain chemical evolution models and
the nucleosynthesis of the r- and s-processes.

\section{Conclusions}

We have developed inhomogeneous chemical evolution models for the
Milky Way halo. We adopted different hypotheses  for the site
  of the r-process, and also included an early production of
 the s-process by fast-rotating massive stars (spinstars). We compared
our predictions for Ba, Y, Sr, and Eu with the abundance patterns of
very metal-poor stars for these elements. Our main conclusions can be
summarized as follows:

\begin{itemize}

\item Independently of the r-process scenarios adopted, the spinstars
  production of s-process is needed to reproduce the spread in the light
   neutron capture elements (Sr and Y) over  the heavy neutron capture elements (Ba and Eu).

\item Our two best models based on two different r-process scenarios when
coupled with the spinstars s-process production are able to reproduce
 the observational data for a set of four neutron capture elements (Sr, Y, Ba, and Eu).

\item The r-process scenarios are not clearly distinguishable with the
  present data for the Galactic halo. Both scenarios reproduce the
  fraction of r-process-rich stars fairly well. The abundance
  measurements of [Eu/Fe] at intermediate metallicity
  $-$2$<$[Fe/H]$<-$1 tend to favor the MRD scenario. More data at this
  intermediate metallicity might provide an important constraint.

\item We predict the contribution of spinstars to be more pronounced 
in specific zones in the chemical abundance ratios, for example,
at high [Sr/Ba] at [Fe/H]$<-$2 or at low [Ba/Fe] for the same metallicity range.
In these zones the stars are expected to present an s-process signature in the Ba isotopic 
ratios, leading to low odd isotopic ratios possibly measurable by the hyperfine splitting
of the Ba lines (in particular, the line at 455.4 nm).

\item The change of the rate of MRD SNe produces differences in the
  predictions of the model that can be  distinguished using the
  distribution functions. This comparison confirms that the rate we have
  chosen produces results that better agree with the observational data.
  There is also a caveat, since we can conclude this only by assuming no
  bias in the observational data depending on the abundance of neutron
  capture elements; so not only a greater number of measured
  stars but also observational data without (or with known) bias are
  needed to proceed a step forward in the understanding of the
  sources of the r-process.

\end{itemize}

Finally, we would like to point out that a possible  method to distinguish
between these two scenarios might be to apply these same
r-process nucleosynthesis prescriptions to another Galactic component.
A different star formation history provides new constraints because a
fast evolution of the metallicity can help to enhance the differences
in the timescales between the two scenarios considered here. We will
study this possibility in a future work by comparing our model predictions
for the halo and the bulge.

\begin{acknowledgements} 

  We acknowledge the heroic efforts of all the very metal-poor star
  hunters as well as those of all the stellar modelers without whom this
  work would not have been possible. We are grateful to P. Creasey for
  carefully reading the manuscript.

\end{acknowledgements}

\bibliographystyle{aa}
\bibliography{spectro}
\clearpage
\begin{appendix}
\section{}
\begin{figure}[ht!]
\begin{minipage}{185mm}
\begin{center}

\includegraphics[width=185mm]{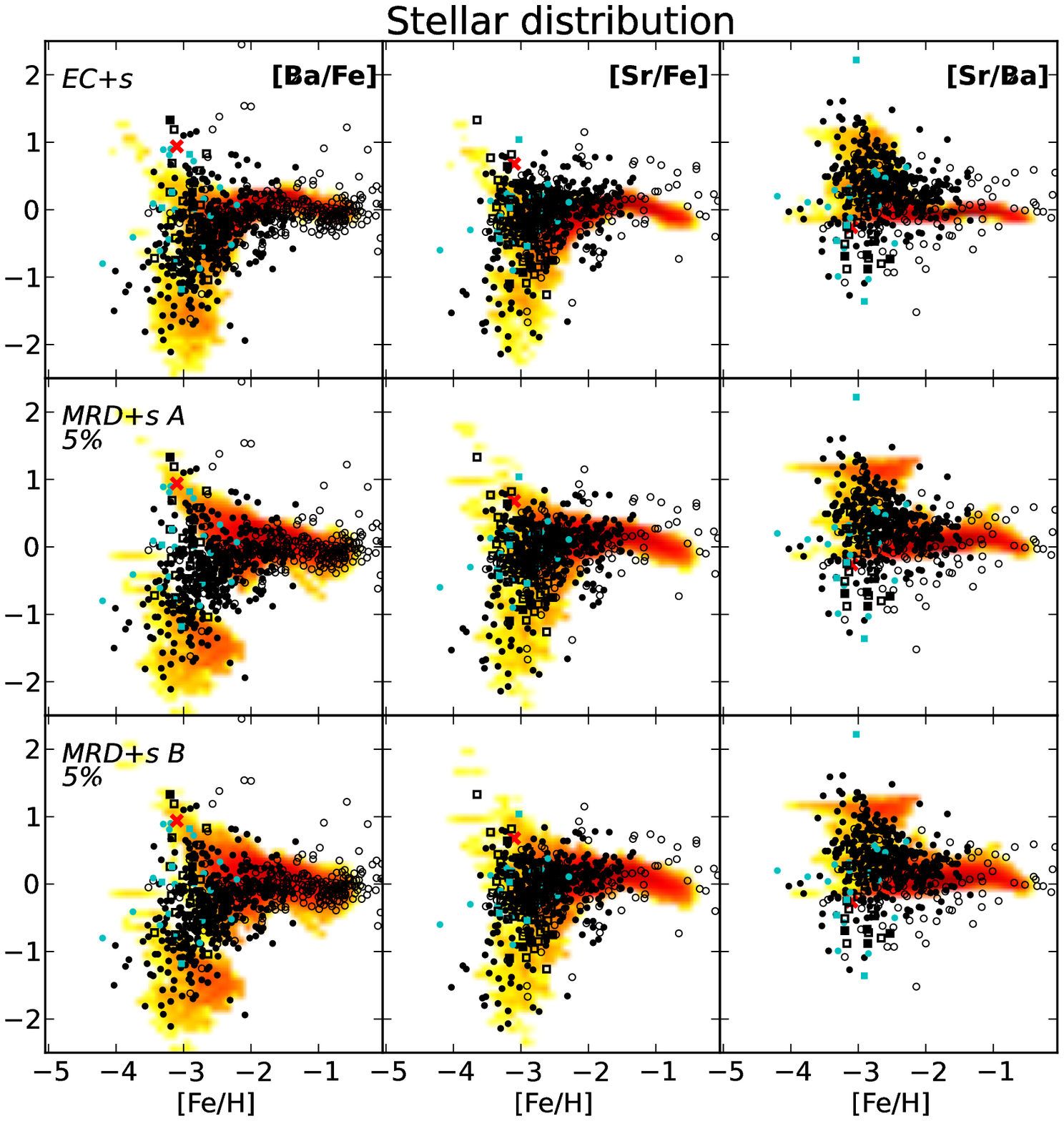}

\caption{As Fig.~\ref{fig1}, but in this case the MRD+s  models
have 5\% as a rate of MRD SNe.}\label{fig7}

\end{center}
\end{minipage}
\end{figure}

\end{appendix}
\end{document}